\documentclass[preprint,12pt]{elsarticle}

\usepackage{epsfig}
\usepackage{amssymb}
\usepackage{amsmath}
 \usepackage{amsthm}

\journal{Optik}

\begin{document}

\begin{frontmatter}

\title{Nonstationary polarization optical forces, considering the influence of dispersion and diffraction} 

\author[label1]{M.-G. Zheleva} 
 \ead{maria.gabriela.zheleva@gmail.com}
\author[label2,label4]{A. Dakova-Mollova} 
 \ead{anelia.dakova@gmail.com}
\author[label3,label4]{V.Slavchev} 
 \ead{valerislavchev@yahoo.com}
\author[label4]{L. Kovachev}
 \ead{lubomirkovach@yahoo.com}
\author[label2]{D. Dakova}
 \ead{ddakova2002@yahoo.com}

\affiliation[label1]{organization={Georgi Nadjakov Institute of Solid State Physics},
            addressline={72 Tsarigradsko Shosse Blvd.}, 
            city={Sofia},
            postcode={1784}, 
            country={Bulgaria}}
\affiliation[label2]{organization={Physics and Technology Faculty, University of Plovdiv "Paisii Hilendarski"},
            addressline={24 Tsar Asen Str.}, 
            city={Plovdiv},
            postcode={4000}, 
            country={Bulgaria}}
\affiliation[label3]{organization={Department of Medical Physics and Biophysics, Medical University - Plovdiv},
            addressline={ Blvd. Vasil Aprilov 15-A}, 
            city={Plovdiv},
            postcode={4002}, 
            country={Bulgaria}}
\affiliation[label4]{organization={Institute of Electronics, Bulgarian Academy of Sciences},
            addressline={72 Tzarigradcko shossee}, 
            city={Sofia},
            postcode={1784}, 
            country={Bulgaria}}

\begin{abstract}
In the present work, the dynamic properties of an attractive longitudinal optical force and the applied potential, due to diffraction and dispersion of ultrashort laser pulses, propagating in air at distances of several diffraction and dispersion lengths, are presented. The results are based on an analytical solution of the linear $3D+1$ paraxial amplitude equation and its application to the evolution in time of the longitudinal optical force. \\
The current research provides valuable guidance for the development and creation of neutral particle laser accelerators with potential applications in the field of laser driven nuclear fusion.
\end{abstract}


\end{frontmatter}


\section{Introduction}
For the first time, a new physical mechanism for trapping neutral atoms, molecules, and particles in a laser beam by a transverse optical force was proposed by Ashkin \cite{Ref1, Ref2}.
Later, in \cite{Ref3}, it was shown that in the pulse regime of propagation, one additional longitudinal polarization force, connected to the time derivative of the Poynting vector, appears. The authors in \cite{Ref4,Ref5} demonstrated that this force draws the particle to the center of the optical pulse, and the inversed intensity profile plays the role of an attractive potential.
The obtained results are approximations of the first order of dispersion at distances considerably smaller than one diffraction length.
But at large distances, the intensity profile of the pulse in the linear regime is changed by diffraction and dispersion.
In the present work, the dynamic properties of the longitudinal optical force and the applied potential due to the influence of dispersion and diffraction are investigated. The results are based on an analytical investigation of the longitudinal optical forces by solving the corresponding linear $3D+1$ paraxial equation for the intensity profile of the laser pulse.

\section{Finding an exact analytical solution to the amplitude equation}
The nonlinear polarization force density of a femtosecond laser pulse propagating in an isotropic medium can be represented as follows \cite{Ref4}:
\begin{align}
\vec{F}=\frac{4\pi}{c^{2}}\frac{d(\chi^{(1)}+\chi^{(3)}\mid\vec{E}\mid^{2})\vec{S}}{dt}, \label{eq1}
\end{align}
where $\mid\vec{E}\mid^{2}=\mid\vec{A}\mid^{2}$ is the square of the electric field modulus, $\chi^{(1)}$ and $\chi^{(3)}$ are the dielectric susceptibility tensors of the medium. The second-order tensor $\chi^{(1)}$ characterizes the linear dielectric susceptibility and is related to the linear polarization of the medium. Nonlinear effects in the optical centrosymmetric medium are mainly due to the nonlinear susceptibility $\chi^{(3)}$. The nonlinear terms in equation (\ref{eq1}) contribute very little when the intensity is below critical for self-focusing and can be approximated as:
\begin{align}
\vec{F}=\frac{4\pi}{c^{2}}\chi^{(1)}\frac{d\vec{S}}{dt}. \label{eq2}
\end{align}
 The magnitude of the Poyting vector $\vec{S}$ characterizes the intensity $I$ of the laser radiation $\vec S=I\vec z$ where the connection between the intensity and the amplitude envelope is well-known. The equation {\ref{eq1}} is transformed in (\ref{eq2}):
\begin{align}
\vec{F}=\frac{2 \chi^{(1)}n_{0}}{c}\frac{\partial \mid A \mid^{2}}{\partial t}. \label{eq3}
\end{align}
The resulting force is proportional to the first derivative of the pulse envelope, while the real force at the level of $\frac{1}{2}$ of the pulse maximum is inversely proportional to the first derivative of the pulse duration. The force described so far is not observed in the CW regime of propagation, while in the femtosecond region it leads to the trapping of particles under the pulse envelope. Under these circumstances, the probability of collision between free atoms and molecules in the air becomes significant.\\
In linear regime of propagation of a three-dimensional laser pulse in isotropic medium, under the influences of the effects of dispersion and diffraction, the spatio-temporal equation describing the evolution of the amplitude function of the optical pulse is as follows \cite{Ref7}:
\begin{align}
i\frac{2k_{0}}{v_{gr}}\frac{\partial A}{\partial t}=\Delta_{\bot} A-\beta \frac{\partial^{2}A}{\partial \xi^{2}},  \label{eq4}
\end{align}
where $ \xi=z-v_{gr}t$ and $\Delta_{\bot}=\frac{\partial^{2}}{\partial x^{2}}+\frac{\partial^{2}}{\partial y^{2}}$ is the transverse Laplace operator. 
The constant $\beta$, characterizing the dispersion, can also be represented as the ratio between the dispersion $z_{dis}$ and the diffraction length $z_{dif}$:
\begin{align}
\beta=k_{0}v_{gr}^{2}\mid k^{\prime \prime}\mid=\frac{z_{dis}}{z_{dif}}, \label{eq5}
\end{align}
where $z_{dis}=\frac{t_{0}^{2}}{\mid k^{\prime \prime} \mid}$ is the dispertion length, $t_{0}$ is the initial time duration of the pulse and $k^{\prime \prime}$ is the group velocity dispersion.\\
The equation is written in Galilean coordinate system. The first term on the right side of the equation characterizes the effect of diffraction and the second one describes the effect of dispersion of the medium.\\
The amplitude equation (\ref{eq4}), describing the evolution of a spatial laser pulse in linear isotropic medium, is a second-order partial differential equation. We find a solution of this equation by using the forward and inverse three-dimensional Fourier transform.
We assume that the initial pulse has a Gaussian shape:
\begin{align}
A(x,y,\xi,0)=A_{0}\exp{(-\frac{x^{2}}{2d_{0}^{2}}-\frac{y^{2}}{2d_{0}^{2}}-\frac{\xi^{2}}{2\xi_{0}^{2}})}, \label{eq6}
\end{align}
where $A_{0}$ is initial amplitude of the laser pulse and $\xi_{0}=v_{gr}t_{0}$ is its initial longitudinal size.
For this purpose, we apply the following algorithm:\\
$1.$ We represent the amplitude function $A(x,y,\xi,t)$ in the form:
\begin{align}
A(x,y,\xi,t) = \frac{1}{(2\pi)^3} \int_{-\infty}^{+\infty} \bigg[\int_{-\infty}^{+\infty} \bigg( \int_{-\infty}^{+\infty} \hat{A}(k_x,k_y,k_{\xi},t) e^{i k_{x} x} e^{i k_{y} y} e^{i k_{\xi} \xi} dk_{\xi} \bigg)dk_{y}\bigg] dk_{x}, \label{eq7}
\end{align}
where $\hat A (k_{x},k_{y},k_{\xi},t)$ is the Fourier transform of the amplitude function in $k$-space. The following transition is used:
\begin{align}
\vec{r}=(x,y,\xi) \rightarrow \vec{k}, \vec{k}=\frac{2 \pi}{\lambda}\vec{e}=(k_{x},k_{y},k_{\xi}),
x \rightarrow k_{x}, y \rightarrow k_{y}, \xi \rightarrow k_{\xi}, \label{eq8}
\end{align}
where $\vec k$ is the wave vector, whose magnitude is the wave number $k=\frac{2\pi}{\lambda}$, and $\vec{e}$ is the unit vector.\\
$2.$ Finding a solution for the Fourier transform of the amplitude function. In this case, we find the corresponding derivatives and substitute the resulting expressions into equation (\ref{eq4}). After a couple of transformations, we obtain the equation for the Fourier transform of the amplitude function:
\begin{equation}
i\frac{2k_{0}}{v_{gr}}\frac{\partial \hat{A}}{\partial t}+\hat{A} (k_{x}^{2}+k_{y}^{2}-\beta k_{\xi}^{2})=0. \label{eq9}
\end{equation}
Equation (\ref{eq9}) is a first-order ordinary differential equation with separable variables. We solve it by separating the variables and then integrating. After couple of transformations, we find the following exact analytical solution for the Fourier transform of the amplitude function in $k$-space:
\begin{equation}
\hat{A}(k_{x},k_{y},k_{\xi},t) = \hat{A}(0,k_{x},k_{y},k_{\xi}) \exp \bigg (\frac{i v_{gr}}{2k_{0}(k_{x}^{2}+k_{y}^{2}-\beta k_{\xi}^{2}t)}\bigg). \label{eq10}
\end{equation}
We have assumed that the initial pulse has a Gaussian shape (\ref{eq6}). Its Fourier transform is:
\begin{equation}
\hat{A}(0,k_{x},k_{y},k_{\xi}) = (2\pi)^{\frac{3}{2}} d_{0}^{2} \xi_{0} A_{0} \exp\bigg[- \frac{(k_{x}^{2} + k_{y}^{2}) d_{0}^{2} + k_{\xi}^{2} \xi_{0}^{2}}{2}\bigg]. \label{eq11}
\end{equation}
We substitute (\ref{eq11}) into (\ref{eq10}). Thus, we determine the solution for the Fourier transform of the amplitude function in $k$-space in its final form.\\
$3.$ Finding an exact analytical solution for the amplitude function $A(x,y,\xi,t)$. As a result, we substitute (\ref{eq10}) and (\ref{eq11}) into (\ref{eq8}). Thus, we obtain:
\begin{align}
A(x,y,\xi,t) = \frac{d_{0}^{2} \xi_{0} A_{0}}{(2\pi)^{\frac{3}{2}}}\int_{-\infty}^{+\infty} e^{-\frac{k_{x}^{2}}{2} \big( d_{0}^{2} - \frac{i v_{gr}t}{k_{0}} \big)} e^{i k_{x} x} dk_{x}\times \\ \nonumber
\int_{-\infty}^{+\infty} e^{-\frac{k_{y}^{2}}{2} \big( d_{0}^{2} - \frac{i v_{gr}t}{k_{0}} \big)} e^{i k_{y} y} dk_{y}\int_{-\infty}^{+\infty} e^{-\frac{k_{\xi}^{2}}{2} \big( \xi_{0}^{2} + \frac{i\beta v_{gr}t}{k_{0}} \big)} e^{i k_{\xi} \xi} dk_{\xi}.\label{eq12} 
\end{align}
After a number of transformations, we obtain the solution in Galilean coordinate system:
\begin{equation}
A(x,y,\xi,t) = A_{0} d_{0}^{2} \xi_{0} \frac{1}{(d_{0}^{2} - \frac{i v_{gr}t}{k_{0}}) \sqrt{\xi_{0}^{2} + \frac{i\beta v_{gr}t}{k_{0}}}}\exp \bigg( -\frac{x^{2} + y^{2}}{2(d_{0}^{2} - \frac{i v_{gr}t}{k_{0}})} - \frac{\xi^{2}}{2(\xi_{0}^{2} + \frac{i\beta v_{gr}t}{k_{0}})} \bigg). \label{eq13} 
\end{equation}
 It takes into account only the effects of dispersion and diffraction. \cite{Ref8}\\
To present the solution in a clearer and more compact form, having in mind (\ref{eq5}), we transform the multipliers:
\begin{equation}
d_{0}^{2} - \frac{i v_{gr}t}{k_{0}} = d_{0}^{2} \bigg(1 - \frac{i v_{gr}t}{z_{dif}} \bigg), \label{eq14} 
\end{equation}
\begin{equation}
\xi_{0}^{2} + \frac{i\beta v_{gr}t}{k_{0}} = \xi_{0}^{2} \bigg(1 + \frac{i v_{gr}t}{z_{dis}} \bigg). \label{eq15} 
\end{equation}
Thus, we obtain the following analytical solution (\ref{eq6}) of the spatio-temporal paraxial equation (\ref{eq4}):
\begin{equation}
A(x,y,\xi,t) = A_{0} \frac{1}{\bigg(1 - \frac{i v_{gr}t}{z_{dif}} \bigg) \sqrt{1 + \frac{i v_{gr}t}{z_{dis}}}}\exp \bigg( -\frac{{x}^{2} + {y}^{2}}{2 d_{0}^{2} (1 - \frac{i v_{gr}t}{z_{dif}})} - \frac{\tau^{2}}{2 t_{0}^{2} (1 + \frac{i v_{gr}t}{z_{dis}})} \bigg). \label{eq16} 
\end{equation}
We find the square of the modulus of the amplitude function, characterizing the pulse intensity:
\begin{equation}
\mid A \mid^{2} =\frac{A_{0}^{2}}{(1 + \frac{(v_{gr}t)^{2}}{z^{2}_{dif}}) \sqrt{1 + \frac{(v_{gr}t)^{2}}{z^{2}_{dis}}}} \exp \bigg( -\frac{{x}^{2} + {y}^{2}}{d_{0}^{2} (1 + \frac{(v_{gr}t)^{2}}{z^{2}_{dif}})} - \frac{\xi^{2}}{\xi_{0}^{2} (1 + \frac{(v_{gr}t)^{2}}{z^{2}_{dis}})} \bigg) . \label{eq17} 
\end{equation}
Comparing this expression with formula (\ref{eq6}), describing the shape of an initial Gaussian pulse, it becomes clear that under the influence of dispersion and diffraction, the amplitude, longitudinal size and duration of the pulse change. \cite{Ref8}
From the expression (\ref{eq16}), in the case of $z = z_{dis}=z_{dif}$, it can be easily seen that at distance of one dispersion (or diffraction) length, the initial intensity decreases by a factor of $(\frac{1}{2\sqrt{2}})$, and the longitudinal size and duration of the pulse increase by $(2\sqrt{2})$ times. In this particular case for the initial amplitude $A_{0}=1$, the intensity of the pulse at one diffraction length is $0.35355$. \cite{Ref6,Ref8}\\
Here it is important to note that for solids, the group velocity dispersion $(k^{\prime \prime})$ is several orders of magnitude larger than that in air and gaseous media. In the femtosecond range (for light bullets) the conditions for equalizing the dispersion and diffraction lengths exist \cite{Ref4,Ref6}:
\begin{equation}
z_{dis}=z_{dif} \rightarrow \frac{t_{0}^{2}}{\mid k^{\prime \prime} \mid} =k_{0}d_{0}^{2} \rightarrow \frac{z_{0}^{2}}{\beta}=d_{0}^{2} \rightarrow \beta=\delta^{2}=1. \label{eq18} 
\end{equation}
 Figure 1 presents graphs of the trapping potential and the optical force of initial Gaussian pulse.
\begin{figure}[t]
  \centering
  \includegraphics[width=0.45\textwidth]{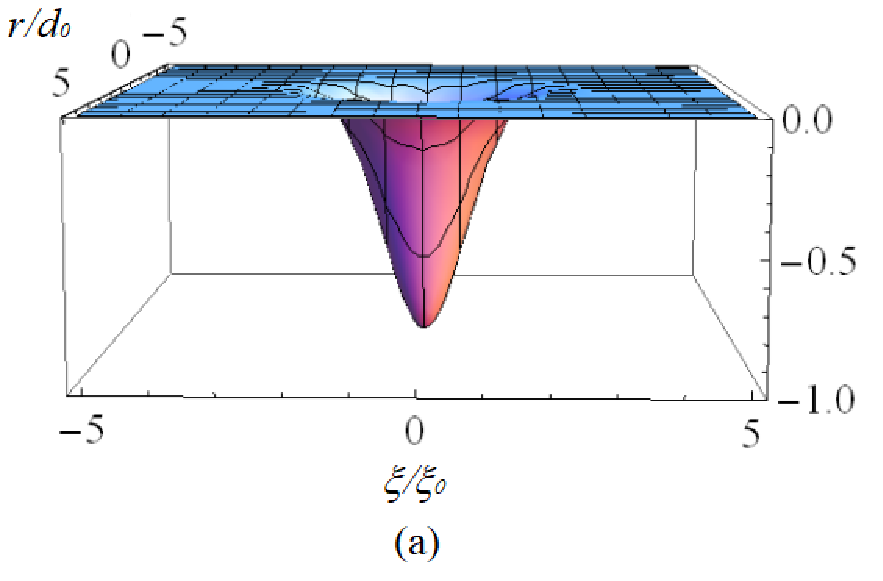}
  \includegraphics[width=0.45\textwidth]{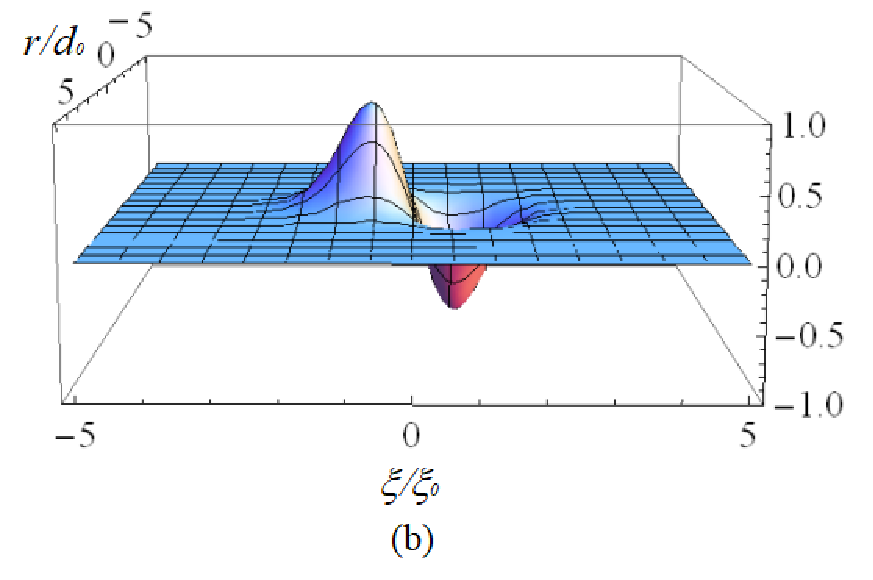}
  \caption{(a) Trapping potential of initial Gaussian pulse for $t=0$, (b) Optical force of initial Gaussian pulse for $t=0$.}\label{fig1}
\end{figure}
In Galilean coordinate system the square of the modulus of the amplitude function takes the form:
\begin{equation}
\mid A \mid^{2} =\frac{A_{0}^{2}}{(1 + \frac{v_{gr}^{2}t^{2}}{z^{2}_{dif}}) \sqrt{1 + \frac{v_{gr}^{2}t^{2}}{z^{2}_{dis}}}} \exp \bigg( -\frac{{x}^{2} + {y}^{2}}{d_{0}^{2} (1 + \frac{v_{gr}^{2}t^{2}}{z^{2}_{dif}})} - \frac{\xi^{2}}{\xi_{0}^{2} (1 + \frac{v_{gr}^{2}t^{2}}{z^{2}_{dis}})} \bigg). \label{eq19} 
\end{equation}
As we already mentioned before, it is considered the linear regime of propagation of optical pulses. Therefore, the nonlinear effects due to $\chi^{(3)}$ are neglected. In this case, the polarization force density is given by the expression:
\begin{equation}
\vec{F}(x,y,\xi,t)=\frac{2n_{0}\chi^{(1)}v_{gr}}{c}\frac{\partial \mid A \mid^{2}}{\partial \xi}\vec{\xi_{0}}. \label{eq20} 
\end{equation}
We determine that:
\begin{equation}
\frac{\partial \mid A \mid^{2}}{\partial \xi}=-\frac{2\mid A \mid^{2}}{\xi_{0}^{2}(1+\frac{v_{gr}^{2}t^{2}}{z_{dis}^{2}})}\xi. \label{eq21} 
\end{equation}
Having in mind that:
\begin{equation}
\vec{ F}(x,y,\xi,t)=-f(x,y,\xi,t)\vec \xi_{0}, \label{eq22} 
\end{equation}
where $\vec \xi_{0}$ is the unit vector.
We substitute (\ref{eq19}) into (\ref{eq20}) and after couple of transformations we obtain an expression for the required force:
\begin{equation}
f(x,y,\xi,t)=-\frac{4n_{0}\chi^{(1)}v_{gr}}{c}\frac{2\mid A \mid^{2}}{\xi_{0}^{2}(1+\frac{v_{gr}^{2}t^{2}}{z_{dis}^{2}})}\xi. \label{eq23} 
\end{equation}
As we can see, the force is negative. Therefore, the front part of the pulse attracts the particles towards its center, while the trailing part pushes them in the same direction.\cite{Ref8}
For ultra-short optical pulses with an initial time duration of $35 fs$, the dispersion length can be assumed to be equal to the diffraction length.
The potential density can be introduced by the expression:
\begin{equation}
dU(x,y,\xi,t)=\vec{F}(x,y,\xi,t)d\vec \xi. \label{eq24} 
\end{equation}
The minus sign indicates that the negative work is considered when the force acts in the opposite direction to the translation vector $ d \vec \xi=d\xi \vec \xi_{0}$. The force depends on the spatial coordinates. The trapping potential can be introduced by integrating (\ref{eq17}) and using the relation $I_{0}=\frac{n_{0}c\mid A \mid^{2}}{(2/\pi)}$: \cite{Ref4, Ref8}
\begin{equation}
dU(x,y,\xi,t)=- f d\xi, \label{eq25} 
\end{equation}
\begin{equation}
U=-\int f d\xi, \label{eq26} 
\end{equation}
\begin{equation}
U(x,y,\xi,t)=-\frac{2n_{0}\chi^{(1)}v_{gr}}{c}\frac{A_{0}^{2}}{(1 + \frac{v_{gr}^{2}t^{2}}{z^{2}_{dif}}) \sqrt{1 + \frac{v_{gr}^{2}t^{2}}{z^{2}_{dis}}}} \exp \bigg( -\frac{{x}^{2} + {y}^{2}}{d_{0}^{2} (1 + \frac{v_{gr}^{2}t^{2}}{z^{2}_{dif}})} - \frac{\xi^{2}}{\xi_{0}^{2} (1 + \frac{v_{gr}^{2}t^{2}}{z^{2}_{dis}})} \bigg). \label{eq27} 
\end{equation}
On figure 2  are presented graphs of the trapping potential and the optical force of a femtosecond laser pulse with $t_{0}=35 fs$ when $z_{dis}=z_{dif}$.
\begin{figure}[t]
  \centering
  \includegraphics[width=0.45\textwidth]{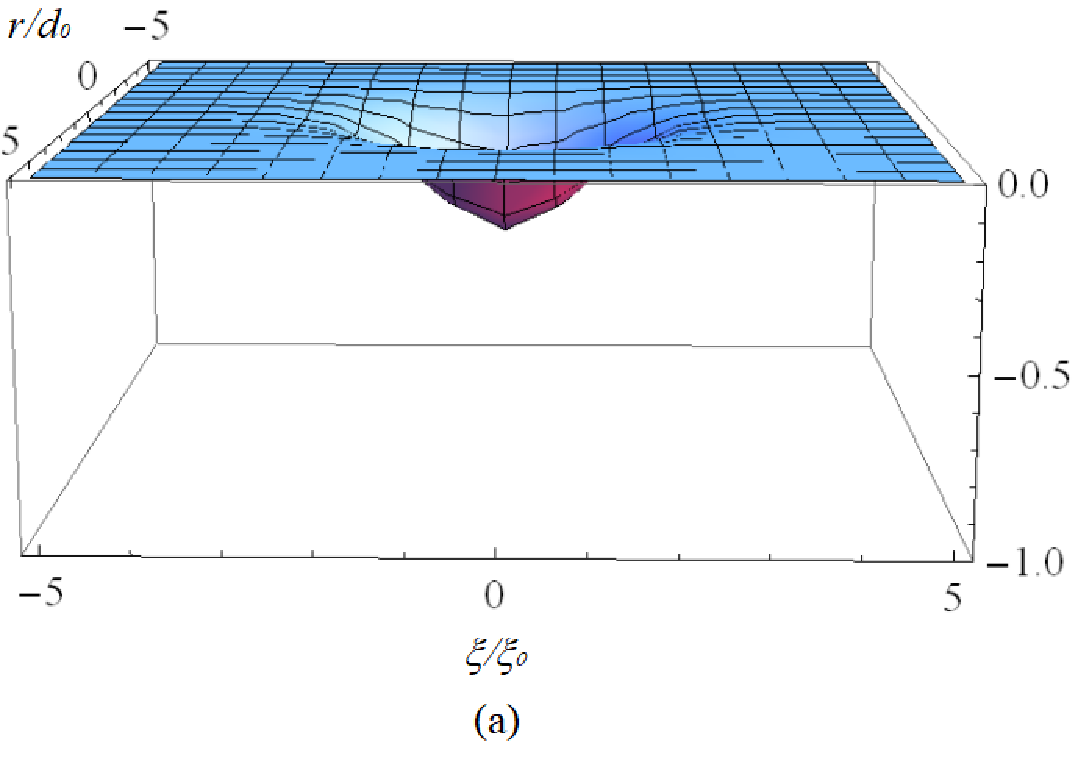}
  \includegraphics[width=0.50\textwidth]{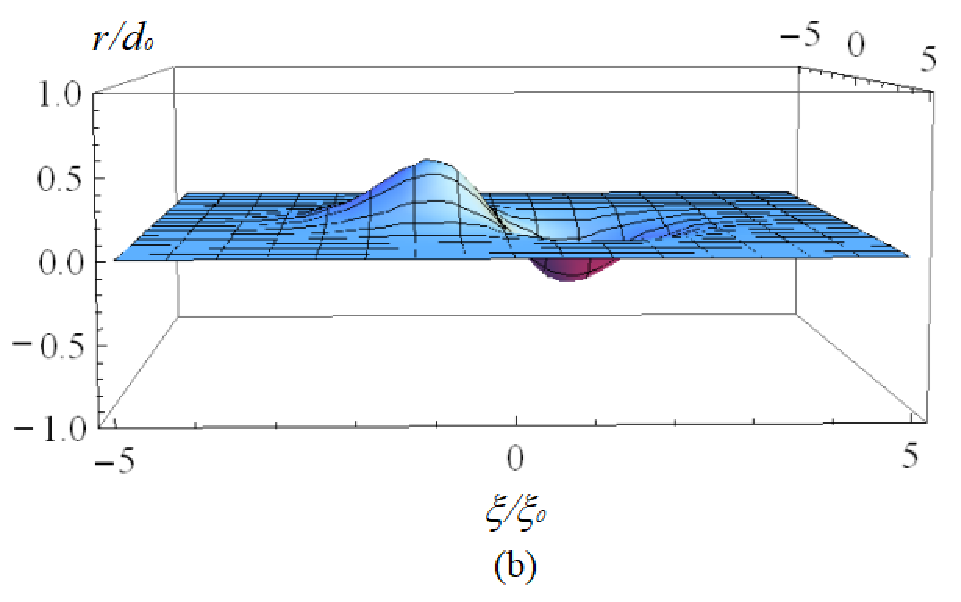}
  \caption{(a) Trapping potential of laser pulse at $z_{dis}=z_{dif}$, (b) Optical force of laser pulse at $z_{dis}=z_{dif}$.}\label{fig2}
\end{figure}

As a gradient force, the optical force is inversely proportional to the longitudinal length or time duration of the laser pulse. Long pulses exhibit relatively small gradients compared to femtosecond pulses, making this force negligible for longer pulses. \cite{Ref6}

\section{Conclusions}
In the current work, the dynamic characteristics of the attractive longitudinal optical force and the applied potential, due to diffraction and dispersion of ultrashort laser pulses at distances of few diffraction and dispersion lengths, are presented. The obtained results are based on analytical solutions of the linear 3D+1 paraxial equation with application to the longitudinal optical force.\\
The study of the effect of these forces on a continuous medium and an ensemble of neutral particles shows that the force applied to an atom is transformed into a volume density force. The optical response of the dielectric media associated with the propagation of the laser pulse is not stationary. The nonlinearity does not add significant amounts to the force and therefore in the diffraction-free regime it can be neglected.\\
The propagation of a laser pulse in the linear regime is characterized by the dispersion and diffraction. Therefore, the amplitude envelope of a linearly polarized optical pulse can be presented up to the second-order dispersion approximation in Galilean coordinates. The resulting force density is negative. This is physically characterized by the fact that the front of the pulse attracts particles towards the centre, while the back also pushes them towards the centre. The minus sign indicates that the negative work is in the opposite direction of the translation vector and that leads to an attractive potential connected to this force.\\
The present investigations may contribute to the creation of neutral particle laser accelerators with potential applications in medicine, power and energy generation.

\section*{CRediT authorship contribution statement}
M.-G. Zheleva: Investigation, Methodology, Writing - original draft; A. Dakova-Mollova: Investigation, Validation, Writing – original draft; V. Slavchev: Visualization, Validation; L. Kovachev: Supervision, Formal analysis, Conceptualization; D. Dakova: Methodology, Investigation, Validation

\section*{Declaration of Competing Interest}
The authors declare that they have no known competing financial interests or personal relationships that could have appeared to influence the work reported in this paper.

\section*{Acknowledgements} The present work is funded by the Bulgarian Ministry of Education and Science under contract $\rm N^{\underline o}$D01-351/2023 and The Bulgarian National Science Fund under contract KP-06-H78/3.

\section*{Data availability}
No data was used for the research described in the article.

\end{document}